%% file: sdnddos.tex
\documentclass[journal]{IEEEtran}
\usepackage{graphicx}
\usepackage{caption}
\usepackage{hyperref}
\ifCLASSINFOpdf
\else
\fi

\usepackage[english]{babel}
\usepackage{url}

\begin{document}
\title{The role of Blockchain in DDoS attacks mitigation : techniques, open challenges and future directions}

\author{Rajasekhar~Chaganti,
         Bharat~Bhushan, Vinayakumar Ravi \thanks{Rajasekhar Chaganti was with the Department
of Computer Science, UTSA, San Antonio,
TX, 78256 USA e-mail: raj.chaganti2@gmail.com}
% \thanks{Bharat Bhushan is with the Department of Computer Science and Engineering,School of Engineering and Technology (SET), Sharda University, India. e-mail: bharat_bhushan1989@yahoo.com}
\thanks{Vinayakumar Ravi is with the Center for Artificial Intelligence, Prince Mohammad Bin Fahd University, Khobar, Saudi Arabia. e-mail:(vravi@pmu.edu.sa)}}

\maketitle

\begin{abstract}
With the proliferation of new technologies such as Internet of Things (IOT) and Software-Defined Networking (SDN) in the recent years, the distributed denial of service (DDoS) attack vector has broadened and opened new opportunities for more sophisticated DDoS attacks on the targeted victims. The new attack vector includes unsecured and vulnerable IoT devices connected to the internet, denial of service vulnerabilities like southbound channel saturation in the SDN architecture. Given the high-volume and pervasive nature of these attacks, it is beneficial for stakeholders to collaborate in detecting and mitigating the denial of service attacks in a timely manner. The blockchain technology is considered to improve the security aspects owing to the decentralized design, secured distributed storage and privacy. A thorough exploration and classification of blockchain techniques used for DDoS attack mitigation is not explored in the prior art. This paper reviews and categorizes the existed state-of-the-art DDoS mitigation solutions based on  blockchain technology. The DDoS mitigation techniques are classified based on the solution deployment location i.e. network based, near attacker location, near victim location and hybrid solutions in the network architecture with emphasis on the IoT and SDN architectures. Additionally, based on our study, the research challenges and future directions to implement the blockchain based DDoS mitigation solutions are discussed. We believe that this paper could serve as a starting point and reference resource for future researchers working on denial of service attacks detection and mitigation using blockchain technology. 

% In addition, the DDoS attack detection and mitigation techniques targeting blockchain peer-to-peer networks or exchanges including the attack type and classifications are presented. 

\end{abstract}


% \cite{Ahmed2018}

% \fi
\IEEEpeerreviewmaketitle

\input{1introduction}
\input{2background}
\input{4ddosusingblockchain}
\input{6challenges}
\input{5futuredirection}
\input{7conclusion}

{
\bibliographystyle{unsrt}
\bibliography{references}
}

% or


% \appendices
% \section{Proof}

% % you can choose not to have a title for an appendix
% % if you want by leaving the argument blank
% \section{}

% % use section* for acknowledgment
% \section*{Acknowledgment}

\ifCLASSOPTIONcaptionsoff
  \newpage
\fi

\end{document}

%% file: 1introduction.tex
\section{Introduction}
\label{sec:intro}
In the recent years, distributed denial of service (DDoS) attacks has been growing and always seen the upward trend \cite{RealSecurity2019EvolutionSecurity}. Work from home and increased use of cloud technologies owing to the Covid pandemic in the first quarter of 2020 has increased the volume and intensity of DDoS attacks in 2020. For example, launching various amplification and UDP-based attacks to flood target networks increased 570 percent for the second quarter of 2020 in comparison with the previous year for the same time period \cite{Nexusguard2020DDoSQ2}; the traditional threshold-based mitigation methods are insufficient to detect these attacks and the machine learning models are able to accurately detect as long as the attack pattern follows the trained data model and if any new attack pattern can easily evade these models \cite{Nexusguard2020DDoSQ2}. Although the DDoS attack vectors existed for years and many solutions proposed for handling the attacks, it is still an important problem to be addressed as the new technologies increases the attack surface and exploitable vulnerabilities. 

\begin{figure}[!h]	
\centering
\includegraphics[width=8 cm, height=6cm]{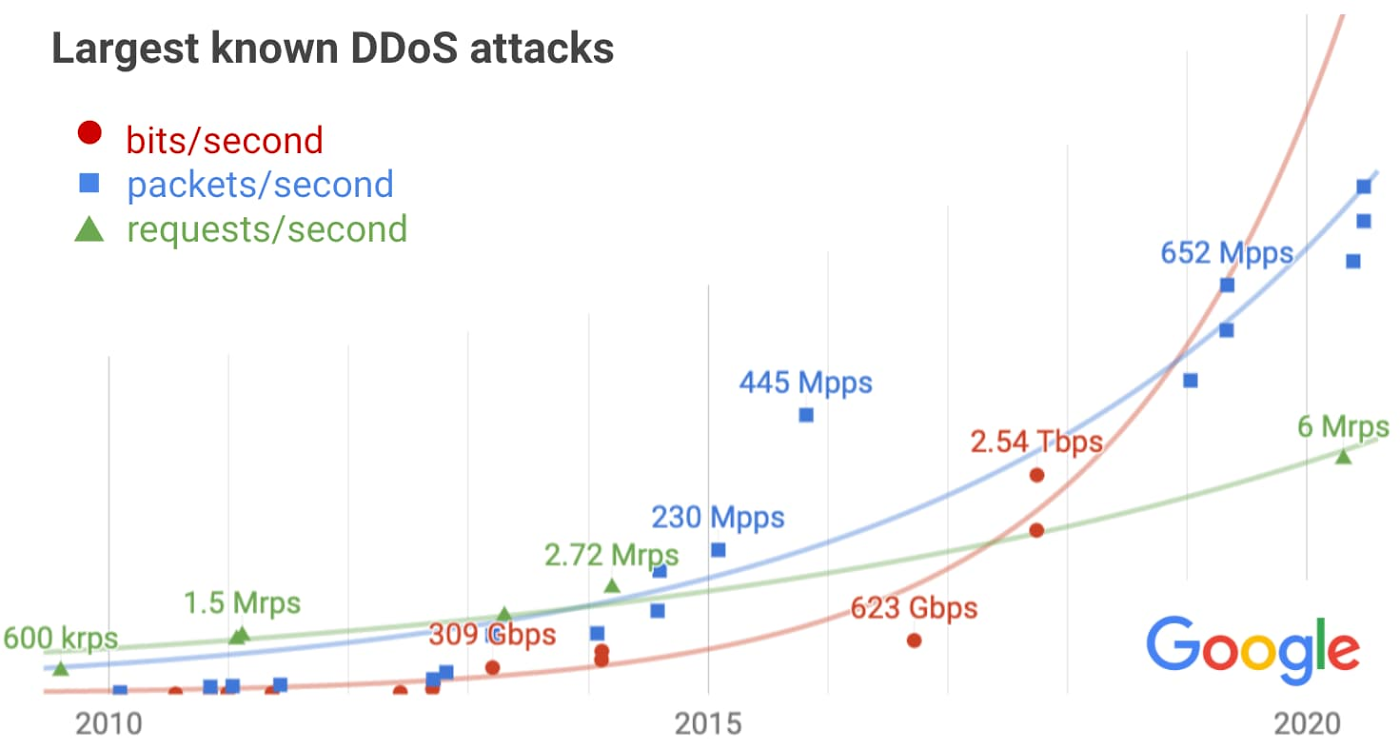}
%\captionsetup{justification=centering}
\caption{DDoS attack rate growth trend in the last decade \cite{Menscher2020IdentifyingBlog}.\label{Attackgrowth1}}
\end{figure}

As the number of devices connected to the internet increases and new network protocol vulnerabilities are uncovered, e.g., the UDP Memcached vulnerability \cite{Newman2018AWIRED}, DDoS attack rates have increased exponentially over the last decade, as shown in Figure \ref{Attackgrowth1}. A nominal enterprise organization may not be able to effectively handle or mitigate the current terabit rate sized attacks, and it's already late to bring up the network Operators and internet service providers to react and mitigate DDoS attacks when attackers target these enterprises. However, as mentioned in Table \ref{AttacksHistory}, we can see that the cloud service providing organizations like Amazon Web Services (AWS) and Google Cloud Platform (GCP) were handled approximately more than 2 Tbps attack rate at the edge level and served the public cloud application customers with no performance or service impact in the last two years. In 2016, the IOT devices  such as routers and cameras connected to the internet were compromised, and attack code deployed to launch mirai bot reflection attacks to generate attack traffic rates in excess of 1 Tbps targeting  DYN (a dynamic DNS service provider), OVH (cloud service provider), and security blogger Brian Krebs’s website \cite{BrianKrebs2016KrebsOnSecuritySecurity} \cite{Kennedy2016OVHWebcams}\cite{Vaughan-Nichols2016TheZDNet}. 

The emerging technologies such as cloud Computing, Internet of Things (IoT), Software Defined Networking (SDN) change the internet network architecture and offers new opportunities for the attackers finding the loopholes and perform Denial of service attacks. The challenge of large-scale DDoS attacks is to mitigate them within a short span of time and avoid the loss of business and reputation for the enterprise organizations involved in the attack. Therefore, a rapid coordination and response required between the stakeholders like network operators, edge protection providers, Internet service providers, impacted organizations, third party DDoS mitigation services etc. Authenticating and establishing trust among the parties involved  is essential to execute the legitimate actions for stopping the attacks.

A blockchain is a distributed ledger that can record the transactions in an efficient and permanent way. It is managed by peer-to-peer (P2P) network nodes with standard protocols designed for internode communication to approve the transaction records and validate the blocks. Owing to the inherent security by design and unalterable transaction records in the chain of blocks, a blockchain can be used for many applications including finance, healthcare, supply chain, cryptocurrency, cybersecurity, smart contacts in particular validating the identity, providing the user anonymity \cite{Zheng2019}\cite{BlockchainWikipedia}. The blockchain utility for cybersecurity application has been growing with demand to build secured systems and applications. The decentralized consortium blockchain implementation for industrial IoT \cite{Li2018} \cite{Wan2019}, credit based consensus mechanism for approving the transactions in industrial IoT \cite{Huang2019} and implementing blockchain based data storage and protection mechanism for defending the security attacks in IoT systems \cite{Liang2019} \cite{Li2019} are some of the applications of the blockchain in IoT. Additionally, blockchain is leveraged for security in other areas like secured storage of the data in mobile ad hoc networks \cite{Yazdinejad2020}, decentralized DNS database for DNS attacks mitigation such as cache poisoning attacks \cite{Li2021}, secured data storage in cloud and defend against the keyword guessing attacks \cite{Zhang2020}. Furthermore, based on the blockchain exhibiting security properties, we could see that the potential to utilize the blockchain for security threat information sharing among the key stakeholders. 

%\begin{paracol}{2}
%\linenumbers
%\switchcolumn

Recently, a few researchers proposed blockchain based solutions for threat information sharing like malicious IP address for blocklist, identifying the IOT bots in the network at the network gateway level, enabling content distribution network (CDN) nodes near the victim using private blockchain when denial of service is identified, security operating center threat sharing to users accessed in private blockchain is investigated in several recent works \cite{Rodrigues2017EnablingBloSS} \cite{Kim2018DDoSBlockchain} \cite{Badruddoja2020IntegratingSensors} \cite{Yeh2020SOChain:Blockchain} \cite{Tariq2019}. But there is a knowledge gap between network security experts, who aim to mitigate  DDoS attacks in real time and blockchain experts, who develop decentralized applications but may not be experts in network attacks. Our prior art research shows that there is no significant work on investigating blockchain’s role to mitigate the DDoS attacks. Therefore, we believe that there is a need for a systematic thorough review of the blockchain technology to handle the denial of service attacks. In addition, the blockchain based solutions are categorized based on the DDoS mitigation deployment location in internet. To the end,
the main contributions of this paper are as follows:
\begin{itemize}
    \item We performed systematic review and classification of the role of  blockchain technology in DDoS attack detection and blockchain based DDoS mitigation solutions. 
    \item We discussed the open challenges and future directions to implement and propose new solutions for handling DDoS attacks using blockchain.
    \item We categorized and described the existing blockchain related DDoS solutions based on the solution deployment location in the internet architecture.
    \item Our findings show that secured collaboration among the stakeholders to share the DDoS threat indicators with blockchain is achievable while addressing the limitations.
    % \item Analysis of the DDoS attacks on blockchain ecosystems, which includes the blockchain P2P network as well as blockchain service exchanges.
\end{itemize}
The abbreviations used in the paper are given in Table \ref{Abbreviations}. The remainder of this paper is organized as follows: Section \ref{sec:background} discusses the key concepts such as DDoS attacks, Blockchains and Emerging technology network architecture paradigms and related work in association with our topic in the paper. Section \ref{sec:ddosusingbc} presents the Blockchain based solutions to mitigate the DDoS attacks. Section \ref{sec:openchallenges} presents the current open challenges to utilize the blockchain in the context of DDoS attacks. Section \ref{sec:futuredirection}  depicts the future directions in accordance with advancement with Blockchain technology. Section \ref{sec:conclusion} concludes the paper.

% Section V discusses the Denial of service attack detection and mitigation focusing on the Blockchain P2P network or cryptocurrency platforms. 

% The MDPI table float is called specialtable
\begin{table}[!h] 
\caption{List of Abbreviations used in the paper.\label{Abbreviations}}
%\captionsetup{justification=centering}
%%% \tablesize{} %% You can specify the fontsize here, e.g., \tablesize{\footnotesize}. If commented out \small will be used.
\begin{tabular}{|c|l|}
 \hline 
ACK	& TCP Acknowledgement Flag \\ \hline
AMQP &	Advanced Message Queuing Protocol  \\  \hline
AMP	& Asynchronous Messaging Protocol  \\ \hline
API	& Application Programming interface \\ \hline
AWS	& Amazon Web Services  \\ \hline
AS	& Autonomous System   \\ \hline
BFT	& Byzantine Fault-Tolerant  \\ \hline
BGP	& Border Gateway Protocol \\ \hline
CDN	& content distribution network	 \\ \hline
CoAP &	Constrained Application Protocol  \\ \hline
CIDS &	Collaborative Intrusion Detection System  \\ \hline
CLDAP &	Connection-less Lightweight Directory Access \\ \hline
CPU	& Central processing unit  \\ \hline
DDoS &	Distributed Denial of Service \\ \hline
DNS	& Domain Name System  \\ \hline
DOTS & 	DDoS Open Threat Signaling  \\ \hline
DoS	& Denial of Service	 \\ \hline
DOS & Decentralized Oracle Service \\ \hline
DPOS &	Delegated Proof of Stake  \\ \hline
EVM	& Ethereum Virtual Machine   \\ \hline
GRE	& Generic Routing Encapsulation	  \\ \hline
GCP	& Google Cloud Services 	 \\ \hline
HTTPS &	Hypertext Transfer Protocol Secure  \\ \hline
HTTP &	Hypertext Transfer Protocol	  \\ \hline
ICMP &	Internet Control Message Protocol  \\ \hline
IoT	& Internet of Things	 \\ \hline
IPFS &	InterPlanetary File System 	 \\ \hline
IP	& Internet Protocol  \\ \hline
ISP &	Internet Service provider  \\ \hline
KNN &	k-nearest neighbor  \\ \hline
LSTM &	Long short-term memory \\ \hline
MLP	& Multi-Layer Perceptron \\ \hline
ML	& Machine learning \\ \hline
MQTT &	Message Queuing Telemetry Transport \\ \hline
NDP	& Neighbor Discovery Protocol \\ \hline
NTP	& Network Time Protocol \\ \hline
OF	& Open Flow \\ \hline
PBFT &	Practical Byzantine fault tolerance \\ \hline
PCA	& Principal component analysis  \\ \hline
PoS	 & Proof of Stake  \\ \hline
PoW &	Proof of Work  \\ \hline
PSH	& TCP Push flag  \\ \hline
P2P	& Peer to Peer \\ \hline
P4 &	Programming protocol-independent packet processor  \\ \hline
RAM &	Random-access memory  \\ \hline
SDN	& Software Defined Network  \\ \hline
RNN &	Recurrent neural network  \\ \hline
SMTP &	Simple Mail Transfer Protocol  \\ \hline
SNMP &	Simple Network Management Protocol  \\ \hline
SOC &	Security Operating Center  \\ \hline
SYN &	TCP Synchronization Flag  \\ \hline
TCP &	Transmission Control Protocol  \\ \hline
SVM	& Support Vector Machine  \\ \hline
TX &	Transaction  \\ \hline
UDP  &	User Datagram Protocol  \\ \hline
UTX	& Unspent Transaction Unit  \\ \hline
XMPP &	Extensible Messaging and Presence Protocol	 \\ 
\hline
% \bottomrule
\end{tabular}
\end{table}

%% file: 2background.tex
\section{Key Concepts and Related Work}
\label{sec:background}
In this section, we review DDoS attack types and the solutions proposed to mitigate them, describe the main fundamental and terminology  of blockchain technology, and describe the emerging technologies such as internet of things and software defined networking paradigm. These are essential and play a significant  role in the understanding of recent DDoS attack variants and their mitigation solutions using blockchain.

\subsection{DDoS Attack Types and Known Solutions}
Distributed Denial of Service (DDoS) Attack is a well-known and major concern in cybersecurity area violating the security principle “Availability” of services.  DDoS attack vectors exploit various features of the internet protocols, most of which were designed decades ago when security was not a concern. The relationship between an attacker exploiting the protocol features such as TCP connection setup using 3-way handshake and its victim is asymmetric in nature. DDoS attacks are mainly classified into two categories: bandwidth depletion and resources depletion attacks \cite{Douligeris2003DDoSClassification}. In the former attack, high volumes of traffic that looks legitimate but not intended for communication is directed to a victim. In the latter attack, the victim is inundated with bogus service requests that deplete its resources and prevent it from serving legitimate requests.
Multiple bots (network nodes compromised and controlled by an attacker) are often used to launch DDoS attacks. Direct attacks on a victim typically use flooding in which many packets are sent from multiple bots to the victim; examples include TCP SYN floods, UDP floods, ICMP floods, and HTTP floods \cite{Swami2018SoftwareMechanisms}. 

Another tactic used in DDoS attacks is amplification: the attacker sends requests to network service providers such as Domain Name System (DNS) servers or network time providers (NTP) spoofing victim’s IP address as the source IP address so that the responses, which are typically several times larger than the queries/requests, are sent to the victim and overwhelm the victim’s network and resources. Examples of  amplification attacks include Smurf, Fraggle, SNMP, NTP, DNS amplification \cite{Swami2018SoftwareMechanisms}.
In addition, protocol exploitation attacks like TCP SYN flooding can be performed on the victim infrastructure by taking advantage of TCP connection establishment  mechanism and  sending the flood of TCP SYN packets with no ACK responses to consume the victim machine resources \cite{Srivastava20AMechanisms}. The adversary may also use automated scripts to send  TCP flags ACK, PUSH, RST, FIN packet floods to saturate the communication channel along the victim infrastructure. Another category of DDoS attack are ping of death and land attack. Ping of death attack focused on sending Ping command with packet size greater than maximum packet size 65536 bytes to crash the victim the system. In land attack, An attacker may send forged packets with same sender and destination IP address to target the victim to send the packet to itself forming an infinite loop and crashing the victim machine \cite{Srivastava20AMechanisms}.  A zero-day can vulnerability also be leveraged to compromise the legit machines and successfully lunch the denial of service attack \cite{CatalinCimpanu2020DDoSZDNet}.

Significant research work is done on the detection and mitigation of DDoS attacks  for the last two decades. The proposed mitigation solutions differ in the location and timing of deployment \cite{Zargar2013AAttacks}. The deployment location-based solutions are categorized into four types
\begin{itemize}
\item Source-based defense implemented in the attack source edge routers or source Autonomous Systems.
\item  Destination-based implemented at the victim edge routers or victim AS level.
\item Network-based defense implemented by the ISP and core networks and usually required to respond the attacks at the intermediate network level and \item  Hybrid defense : the combination of the source, destination and network based mechanisms.
\end{itemize}
Although the source-based defenses aim to detect and mitigate the attacks in early stages of the attack, it is very difficult to distinguish the legitimate and malicious DDoS traffic at the source level owing to the use of bots to distribute the attack traffic generation. 

% The MDPI table float is called specialtable
\begin{table*}[!h] 
\captionsetup{justification=centering}
\caption{Major DDoS attacks in the history.\label{AttacksHistory}}
%\resizebox{\textwidth}{!}{%
\begin{tabular}{|p{2.3cm}|p{0.5cm}|p{2.6cm}|p{1.7cm}|p{1cm}|p{1.4cm}|p{2.5cm}|p{2cm}|}  \hline
%\toprule
\textbf{DDoS Attack}	& \textbf{Year} & \textbf{Attack Type} & \textbf{Attack Rate} & \textbf{Duration}   &   \textbf{Amp Ratio} & \textbf{Protocols Involved}   &  \textbf{Impact} \\ \hline
%\midrule
AWS Attack \cite{Amazon2020AWS2020}[18] & 2020 & Reflection Attack & 2.3 Tbps &	3 days & 56 - 70 & UDP, CLDAP	& No \\ \hline
Google Attack \cite{Menscher2020IdentifyingBlog} & 2017 &	Reflection & 2.5 Tbps &	6 months &	6-70 &	CLDAP, DNS, SMTP &	No \\ \hline
Mirai Krebs \cite{BrianKrebs2016KrebsOnSecuritySecurity}  & 2016 & Mirai, TCPSYN, ACK, ACK+PSH & Krebs \- 620Gbps  &	2-7 days & - & TCP, GRE, HTTP	 & Krebs Offline \\ \hline

OVH \cite{Kennedy2016OVHWebcams} &  2016 &  Mirai, TCPSYN, ACK, ACK+PSH & OVH \- 1.1 Tbps & 2-7 days & - & TCP, GRE, HTTP & OVH minimal \\ \hline
Mirai Dyn \cite{Vaughan-Nichols2016TheZDNet} & 2016 &	Mirai, Reflection & 	1.5Tbps	& 1 day &	Up to 100 & DNS 	& Internet Outage \\ \hline
GitHub Attack \cite{Newman2018AWIRED} & 2018 &	Memcached Reflection &	1.35Tbps &	20 min & ~51000 &	UDP &	Service Outage \\ \hline
Six Banks \cite{Constantin2012DDoSCIO} & 2012 &	Brobot &	60 Gbps	&  \~ 2 days &	- &	 HTTP, HTTPS, DNS, TCP &	Web Service Outage \\ \hline
Hongkong Central \cite{RUSSELL2014HongAttack} & 2014 & Brobot,TCP SYN, HTTPS Flood & 500Gbps & - & - & TCP,HTTPS & Minimal \\ \hline
Spamhaus \cite{JaikumarVijayan2013Update:Computerworld} & 2013 & Reflection Attack & 300 Gbps & - &  Up to 100 & DNS,TCP & Offline  \\ \hline
Cloudflare \cite{Thompson2014Record-breakingNetwork} & 2014 & Reflection Attack & 400 Gbps &	- &	Up to 206 &	NTP & No \\ 
\hline
%\bottomrule
\end{tabular}
%%}
\end{table*}

The destination-based defense mechanisms are easier and cheaper to implement since the attack traffic will be concentrated closer to the victim. However, before they are detected; the attack traffic consumes the resources on the paths leading to the victim. The network-based defense solutions detects and mitigate the DDoS attacks at the Autonomous System (AS) or Internet Service Provider (ISP) levels, which are closer to the attack sources. But they incur storage and processing overhead at the network infrastructure level, for example, by the edge or ISP routers, or might need additional DDoS protection devices like middle boxes to process the traffic. Also, the attack detection will be difficult owing to lack of aggregation of traffic destined to the victim. However, attack mitigation in the internet core has the advantage of not passing the traffic till the victim network and preventing congestion of communication channel with attack network traffic as well as saving the victim’s computing and network resources. The hybrid defense approach promises to be more robust since it allows to use the combination of defensive mechanism to defend against DDoS attacks. Furthermore, detection and mitigation can be implemented more efficiently. For instance, the detection can occur at the destination or network level and the mitigation technique can be applied near the  source to effectively handle the DDoS attacks. However, its implementation is more challenging because it requires collaboration and cooperation between different entities to exchange attack information without receiving sufficient incentives for some of the participants like service providers \cite{Zargar2013AAttacks} and there needs to be trust between the stakeholders, given the fact that the service providers are diverse and not easy to trust the entities. 

For descriptions of various DDoS mitigation techniques such as anomaly or signature-based detection, machine learning algorithms to attack detection, scrubbing, rerouting, and filtering/blocking techniques, see Zargar et al. \cite{Zargar2013AAttacks}.

\subsection{Blockchain Technology and Their Types}
A blockchain is a digital, public ledger that records list of transactions and maintains the integrity of the transactions by encrypting, validating and permanently recording transactions \cite{BlockchainBankrate.com}. Blockchain technology has emerged as a potential digital technology disrupting many areas including financial sector, security, data storage, internet of things and more. One of the best known uses of blockchains is the design of cryptocurrencies such as Bitcoin \cite{Nakamoto2009Bitcoin:System,Nakamoto2009Bitcoin:System,2021CryptocurrencyCoinMarketCap}.

\begin{figure}[!h]	
\centering
%\captionsetup{justification=centering}
\includegraphics[width=8cm, height = 8cm]{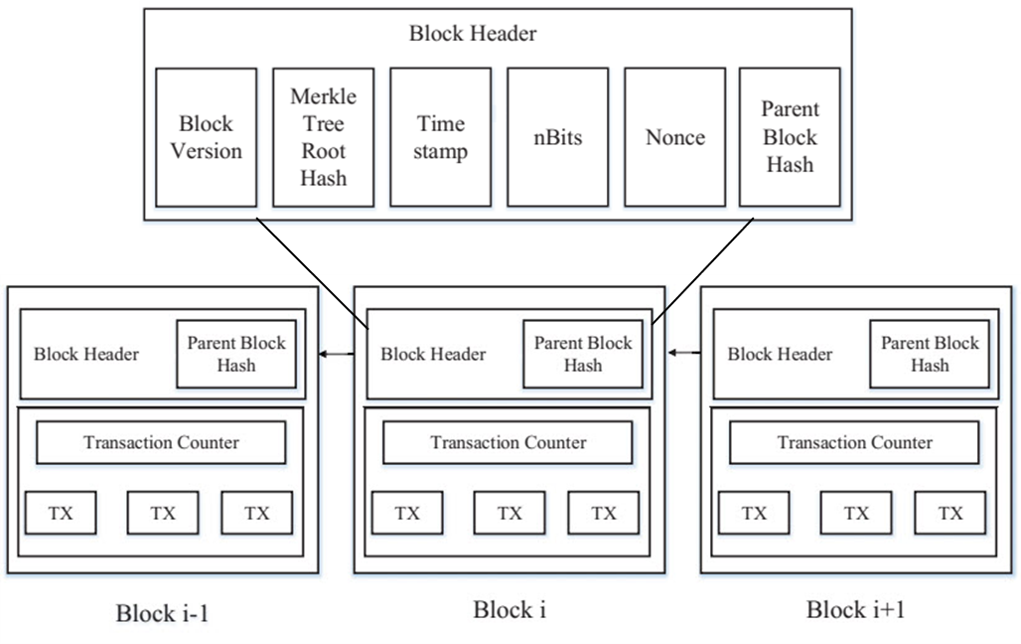}
\centering
\caption{Blockchain Internal Components \label{BCcomponents}}
\end{figure}

A blockchain is typically managed by a peer-to-peer network and uses peer-to-peer protocol such as the Distributed Hash Table (DHT) for internode communication as well as validating new transactions. Figure ~\ref{BCcomponents} illustrates the typical structure of a block: a linked list of blocks with a header block. Each block comprises a set of transactions, a count of the transactions in the block, and a header. The block header includes block version, which tells the current version of block structure, a merkle tree root hash to incorporate the uniqueness of the transaction set in the block by determining the final hash value achieved from all the transactions in the block as well as maintain the integrity between the transactions in the block. Therefore, the transactions secured in a blockchain and cannot be tampered. The block header also contains Timestamp, i.e. the time at which the block is created and it plays an important role in extending a blockchain to record new transactions. There is a special data structure that points to the most recent block in a chain. Using the back pointers other blocks in the chain can be accessed.

\begin{figure}[!h]	
\centering
\captionsetup{justification=centering}
\includegraphics[width=8.4cm, height = 6.7cm]{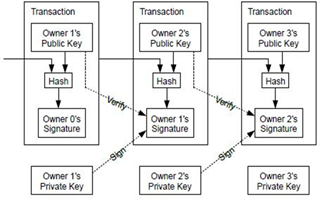}
\centering
\caption{Basic cryptographic operations in blockchain \cite{Golosova2018TheTechnology}. \label{BCcrypto}}
\end{figure}

Blockchain exhibits properties like decentralization, persistency, anonymity, and auditability. The essential property  of anonymity is achieved using asymmetric cryptography like RSA  algorithm and digital signature \cite{Albertorio2018PublicMedium}. Each user has a private and public key pair for applying an asymmetric cryptography algorithm. The hash values obtained from the existing transactions will be utilized to get the digital signature and validate the user’s authenticity. The user validation is a two-step process: signing and verification. Figure ~\ref{BCcrypto} shows the asymmetric cryptography and digital signature calculation steps during the validation process \cite{Golosova2018TheTechnology}. The peer-to-peer blockchain system has no centralized node and uses consensus algorithms, which typically require participating entities to win a computing challenge, to authorize an entity to create the next block of verified transactions and append to the exiting blockchain.

A consensus algorithm, as indicated above, is used to select nodes in peer-to-peer blockchains to add a block of new transactions to the existing blockchain. Some of the widely used algorithms are proof of work (POW), proof of stake (POS), practical Byzantine fault tolerance (PBFT), ripple consensus algorithm and delegated proof of stake (DPOS) \cite{Zheng2017AnTrends}. In POW, used by Bitcoin, every node computes the hash value of the block header and the computed value should be less than the specific value, according to the algorithm. The successfully computed node will be verified by the other nodes and selected as an authorized node to add the transaction in the block; the update is propagated to all other nodes of the blockchain. Computation of the hash value within the constraints requires requires extensive computing, which is called mining. In POS, the users that have more currency can get an authority to add the transactions in the blockchain. So, richer entities will become richer, and, potentially, a few participants dominate the blockchain management and extension; on the other hand, this method does not require extensive computing power, and is likely to more efficient. The consensus algorithm based on PBFT requires that a significant majority of the nodes participating in the blockchain should approve the transaction to be appended in the network and can tolerate 1/3rd of the node failures. The consensus process starts by choosing a primary node to process all the transactions in a block. It is a three-step process i.e.  pre-prepare, prepare and commit; If 2/3rds of the nodes accept the request, then the transaction is appended to the block. Hyperledger's fabric is an example of using PBFT as a consensus mechanism to complete the transactions in the network. In Delegated Proof of Stake(DPOS), the delegated maximum currency stakeholder is chosen for adding the transactions. Some platforms like Tendermint operates on the combination of the algorithms (DPoS+PBFT) \cite{Zheng2017AnTrends}.

With decentralized consensus methods such as POW, branching, in which competing entities may propose different sets of transactions to create a new block and extend a current blockchain, can occur due to the decentralized nature for mining to approve the transaction as well as having a delay to validate the 51\% of the blockchain nodes or participants prior to adding the transaction to blockchain; nBits, which signifies the difficulty level that is being used for miner computations to add the transactions to the block; nonce, which  represents a random number created by the creator of the block and can be used only once; parent block hash, which is a cryptographic hash value of the parent block to maintain the integrity between the two consecutive blocks and maintain the non-tampered chain of blocks \cite{Zheng2017AnTrends} \cite{Zheng2019}.

In general, blockchain platforms are typically classified into three types. Public blockchain, in which the existing transactions can be read by anyone in public and open to join for public. But the transactions cannot be tampered and provide high level security, even though its computation delay is high. Bitcoin is a classic example of public blockchain. Anyone can read the user account balance and the transactions that the user account involved, given the fact that the user bitcoin wallet address is known. In consortium Blockchain, only selected nodes are participated in transactional operations and a good example multiple organization in a particular sector want to use the blockchain for business applications. Each node represents a member from the organization. The consensus process is fast, and only privileged users can read the information from the blockchain. Private Blockchain requires permission to join the network and usually maintained within the organization. The nodes can be the participants from the same organization to share the data within the organization or storing the data records securely and more. The private blockchain usually becomes centralized in nature and the transaction can be tampered if untrustworthy nodes participate in the mining process. The detailed comparison of the blockchain types is described in Table \ref{blockchaintypes}.
% The MDPI table float is called specialtable
\begin{table*}[!h]
% \captionsetup{justification=centering}
\caption{Types of Blockchain and their Properties \cite{Zheng2017AnTrends} \label{blockchaintypes}}
\centering
\begin{tabular}{|l|l|l|l|}
%\toprule
\hline
\textbf{Property}	& \textbf{Public}	& \textbf{Consortium} & \textbf{Private}\\ \hline
%\midrule
Consensus participants  & All mining nodes &	Selected nodes & Nodes within the organization \\ \hline
Efficiency	& Low &	High &	High \\ \hline
Readability & Anyone &	Anyone or restricted members  & Members within the organization \\ \hline
Decentralized	& Yes &	Partial	& No \\ \hline
Consensus authorization  & Permissionless &	Permissioned &	Permissioned \\ \hline
Example & Bitcoin &  R3  & Hyperledger \\ \hline 
Application & Bitcoin currency, voting & Banking, payments & Supply chain, health care, retail \\ \hline
Immutability &	Nearly impossible to tamper & Possibly tampered & Possibly tampered \\ \hline
%\bottomrule
\end{tabular}
\end{table*}

% Revise the next paragraph to discuss why Bitcoin approach does not work DDoS mitigation
Since the existence of the Bitcoin, there are number of coins developed by the blockchain community focusing on specific industry application. Some of the major notable coins are Ethereum, Litecoin and Ripple \cite{REIFF2020TheBitcoin}. The second popular and largest market capitalization cryptocurrency is Ethereum, which works on smart contract functionality. Ethereum has been proposed to address some limitations in Bitcoin scripting language. Ethereum supports the turing complete programming language meaning that we can perform all computations including the loops. This is achieved by smart contracts functionality, which runs cryptographic rules when certain conditions are met. The smart contracts in the nodes are translated into EVM code and then the nodes execute the code to complete the transaction (can be creating a user account, the result of code execution).

There has been a lot of attention on Hyperledger recently owing to the applicability of enterprise standard version blockchain deployment capabilities and known to be rigorously used in academic research community for research activities. Hyperledger is an open source community contributed suite, which comprises tools, frameworks, and libraries for enterprise blockchain application deployments. One of the notable tool is the Hyperledger fabric \cite{Hyperledger/fabric:Privacy.}, a distributed ledger user for developing blockchain applications and can have private blockchain for serving the applications to specific services. The fabric consists of model file, script file, access file and query file and all zipped together to form business network archive. Fabric has a concept called "Chaincode", which is similar to Ethereum smart contract for performing secured blockchain transactions. We can also  include the distributed file storage i.e. Interplanetary File System (IPFS), which store the data and the data can be shared across the nodes in the blockchain. For example, A decentralized web application can be hosted with content stored in IPFS for serving web content to users. Overall, Hyperledger is very useful platform for blockchain technology and have been widely using for developing the applications including DDoS mitigation.

\subsection{Emerging Technology Network Architectures}
Some of the notable recent technologies such as IoT, SDN and cloud computing essentially changed network paradigm. It is important to review these advanced network architectures to study the advanced DDoS attacks exploiting the architecture limitations and propose the new solutions to mitigate these attacks using blockchain technology.

\subsubsection{IOT Architecture}

\begin{figure}[!h]	
\centering
\captionsetup{justification=centering}
\includegraphics[width=8.5cm, height =7cm]{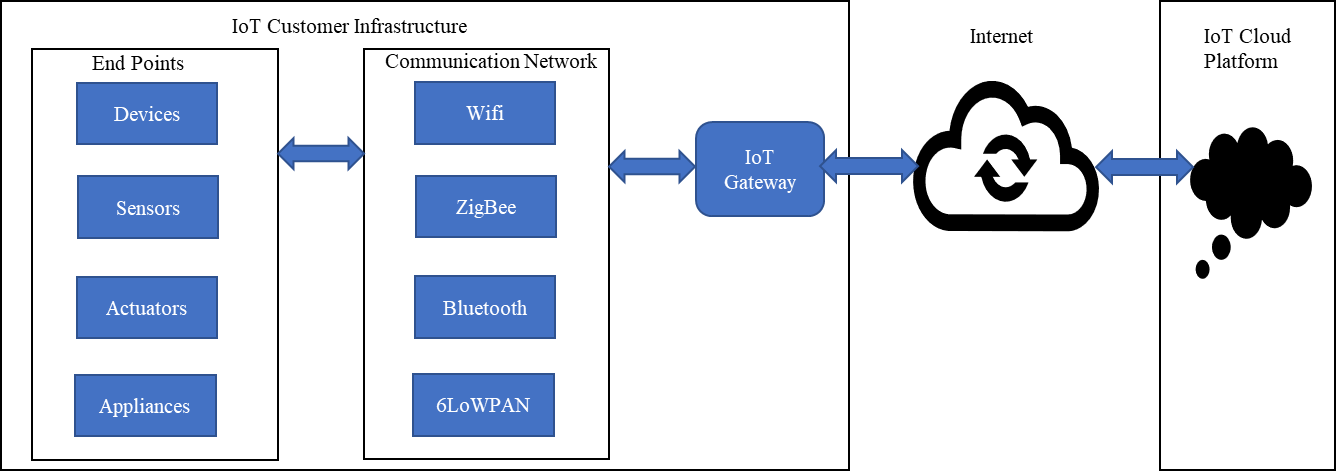} 
\centering
\caption{A typical IoT Architecture. \label{IoTArch}}
\end{figure}

IoT is a system of computing devices including the physical objects with network connectivity to connect to internet and transfer the data over the network with or without requiring the human interaction. The tremendous progress towards smart homes, smart cities, smart transportation, and smart grid applications in recent years shows that rapid advancements in Internet of Things (IOT) technology. Gartner predicted that there will be 65 billion IOT devices connected to the internet by 2025 and the current statistics show that around 31 billion IOT devices deployed and connected to internet \cite{Girad2020TheToday}. Figure ~\ref{IoTArch} depicts a typical IoT architecture with main components. The IoT devices can be sensors, actuators or other appliance installed in home, industry, person body, vehicle, farming platform to monitor or sense the current state or activity and pass the information to the nearest IoT gateway through wireless communication like Bluetooth, Wi-Fi, NFC and ZigBee. The IoT gateways connected to the public internet for sending the information to IoT service provider for data analytics, tracking the status, display in user console etc. Using IoT network protocols such as MQTT, AMP, HTTP and CoAP but not limited. Owing to the limited CPU, memory, and power capabilities of IoT devices and the existence of the multivendor IoT platforms, conventional security solutions are not compatible in IoT environment and securing IoT  devices is challenging.

\subsubsection{SDN Architecture}

\begin{figure}[!h]	
\centering
\captionsetup{justification=centering}
\includegraphics[width=8 cm, height=8cm]{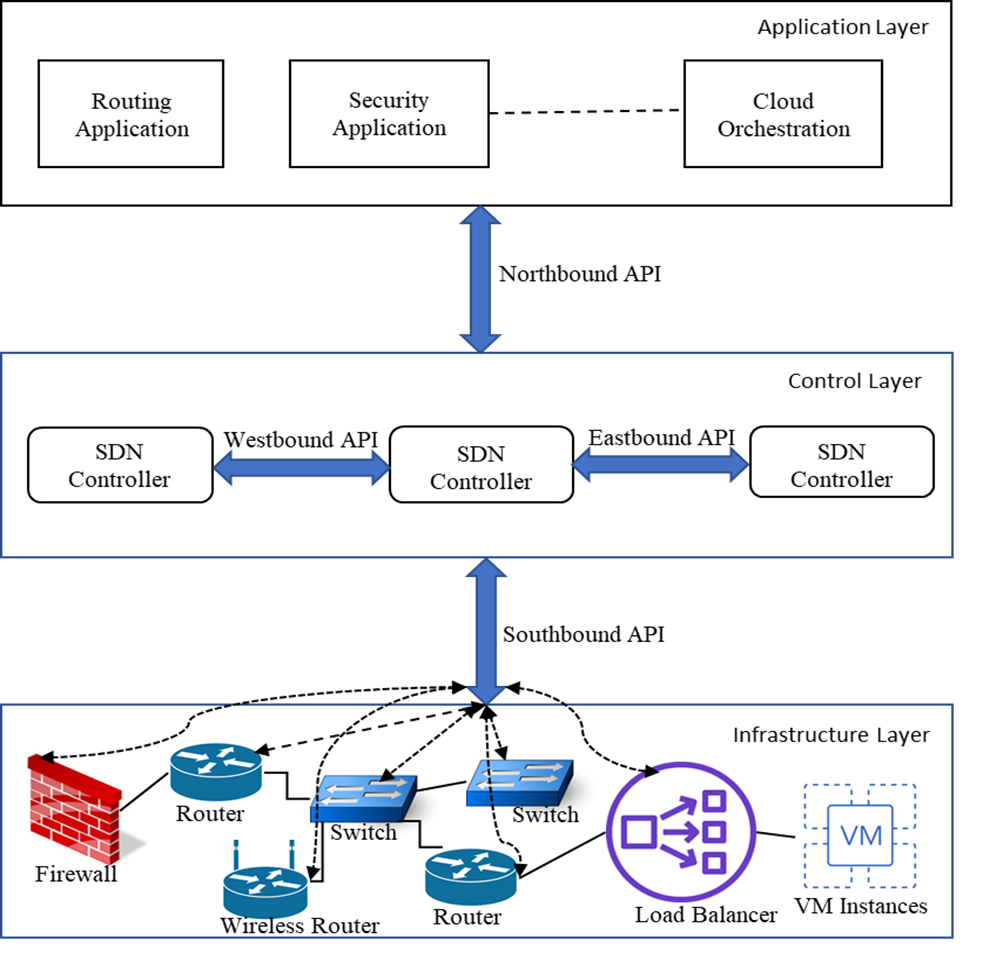} 
\centering
\caption{A typical SDN Architecture} \label{SDNArch}
\end{figure}

Recent advances in wide area networks (WAN) and data center networks are the culmination of the SDN paradigm. SDN enable logically the centralized management of network layer 2 and layer 3 devices  such as Switches and Routers, including the management of wide area networks of the organizations where the network devices located from multiple sites are monitored/controlled using an SDN controller \cite{Govindarajan2014ASolutions}. As depicted in Figure \ref{SDNArch}, the central controller monitors manage all the network device in data plane layer and communicated through southbound API like Openflow standard. A network administrator can develop the applications on top of the control layers to perform network management operations. SDN technology can be used at the  autonomous system level, internet service provider level or data center level for network monitoring and management.  Although SDN provides lot of advantages including programmability, centralized control, and security, it also inherits security vulnerabilities due to the new architecture paradigm. For instance, an adversary may target the controller with TCP SYN flooding attack and other protocol exploitation techniques to saturate the controller and shutdown the whole network \cite{Boppana2020AnalyzingNetworks}. Leveraging the blockchain technology open up new research possibilities to secure the Software defined network itself from malicious denial of service attempts \cite{Huo2020ANetworking} as well as mitigation of the denial of service attacks in conventional networks. 

\subsection{Related Work}

Technologies such as machine learning (ML), blockchain, IoT, and SDN are well suited to improve the security in digital world but also exhibit new security concerns and issues \cite{Lin2017AChallenges} \cite{Ahmad2019AThings} \cite{Muthanna2019} \cite{Wang2019TheSurvey}\cite{Gao2018SecurityNetworks}\cite{Boppana2020AnalyzingNetworks}\cite{Qu2018}\cite{DantasSilva2020}\cite{Dwivedi2019}\cite{Rathore2019}. Some researchers also used combinations of these technologies to address security challenges ranging from malware analysis, DNS Security, to network security as well as privacy issues \cite{Hussain2020MachineChallenges} \cite{daCosta2019InternetApproaches} \cite{Dong2019AEnvironments}\cite{Liu2020ALearning} \cite{Elagin2020}. Our focus in this paper is specific to DDoS-attack detection and mitigation techniques in conventional networks, software defined networks, cloud environments and internet of things \cite{Srivastava20AMechanisms} \cite{Dong2019AEnvironments} \cite{BeslinPajila2020DetectionSurvey} \cite{Sonar2014AThings} \cite{Park2017}. The currently known techniques include machine learning or deep learning (ML/DL) algorithms to classify the attacks, anomaly-based detection, and signature-based detection.

A recent advancement in peer to peer networks with blockchain technology enabled utilization of  decentralized network concepts for multiple application areas like finance, healthcare, real estate, supply chain management, security \cite{BlockchainInsider}. Although blockchain mainly provides the anonymity, privacy and secured data storage in security applications, researchers also explored the applicability of blockchain technology in DDoS attack information sharing, threat intelligence information sharing to quickly respond to the DDoS attacks. Singh et al.  \cite{Singh2020UtilizationAttacks} present a survey of DDoS mitigation techniques using blockchain technology. The authors considered four known blockchain based DDoS mitigation approaches for comparison; highlighted the operation of these mitigation mechanisms and assessed the practical applicability of these implementations \cite{Rodrigues2017AContracts} \cite{BurgerZurich2017CollaborativeBlockchains} \cite{Javaid2018MitigatingBlockchain}\cite{Kataoka2018TrustSDN}. Wani et al. \cite{Wani2021} discussed the prior art distributed denial of service attack mitigation using blockchain by describing the methodology on how the related papers are collected and proposing the taxonomy based on the technologies like artificial intelligence, information sharing capability and blockchain types. However, a comprehensive and systematic review of the state-of-the-art work with classification based on the solution implementation location by leveraging the blockchain technology to detect and mitigate the DDoS attacks in digital world and also detail description of DDoS attacks targeting Blockchain platforms to protect decentralized networks is not covered in the prior art. Our motivation for this work is to bridge the knowledge gap between network security researchers and the blockchain developing community, and enable the researchers to access this article as a reference point to continue the research of using blockchain technology in network security.

%% file: 4ddosusingblockchain.tex
\section{DDoS Attacks Mitigation using Blockchain}
\label{sec:ddosusingbc}
In this section, the existing research works on solving the DDoS attack detection and mitigation problem using blockchain technology is presented and discussed. In addition to blockchain, the role of technologies such as SDN, IoT and ML/DL in addressing DDoS attacks near the attacker domain location, the internet core, or near the victim network domain are reviewed.

% The existing solutions are categorized 

% Table ~\ref{DDoSusingBC} depicts the categorization of the relevant papers with columns representing the paper objective, blockchain technology applied for mitigating the DDoS attacks, the advantage of the authors proposed approach and the limitation of the work if any to make the readers understand the major works in prior art. 
We discuss the existing DDoS mitigation blockchain solutions based on the location of solution deployment in internet architecture.

\subsection{Network level mitigation}
The network level mitigation DDoS mitigation schemes using blockchain technology is deployed at the Internet service provider (ISP) level on the internet, which may be far from attacker or victim location. The Table \ref{DDoSusingBC} illustrates the blockchain key concepts used, technologies involved in the research works proposed for DDoS mitigation using blockchain. We can clearly see that smart contract based Ethereum network is used for implementing the DDoS mitigation solutions for most of the previous contributions, as shown in the Table \ref{DDoSnearNW}. The blockchain access level policy is controlled by the owners to make the transactions accessible for public or private. \\

Tayyab et al. \cite{Tayyab2020ICMPv6} take the approach that each IDS in the network acts as a blockchain node and collaborate with other blockchain IDS nodes to share the attack information like correlated alarms. This decentralized correlated information sharing is used for the detection of ICMP6 based DDoS attacks. Although IDS collaboration improves DDoS attack detection capability, the practical implementation of collaboration can may have difficulties. For example, the IDS vendor interoperability to support the blockchain technology is needed in enterprise environment. Denial of service attacks detection at the IDS level is too late and might already congest the edge network communication channels or the content delivery network communications. \\

The following papers \cite{Yeh2020SOChain:Blockchain} \cite{Pavlidis2020OrchestratingCollaborations} \cite{Abou2019Co-IoT:SDN}\cite{Yeh2019ATechnology}  \cite{Essaid2019ARNN-LSTM} \cite{Yang2019AServices} \cite{Abou2019Co-IoT:SDN} \cite{Kataoka2018TrustSDN} \cite{Rodrigues2017Multi-domainBlockchains} \cite{BurgerZurich2017CollaborativeBlockchains}
\cite{Rodrigues2017AContracts}
\cite{Rodrigues2017EnablingBloSS} focused on utilizing the SDN and blockchain technologies in the autonomous system (AS) level to detect the denial of service attempts and activating the DDoS mitigation mechanisms at the network level. The authors considered the autonomous system consists of SDN architecture, controlled by SDN controller. The core concept in these papers include leveraging the centralized controller application of the SDN to manage how the network devices in the autonomous system should handle the traffic (whitelist/blocklist) originating from malicious IP addresses, which are used to launch the DDoS attacks on the autonomous system. The SDN controller node also acts as a blockchain node running decentralized application like Ethereum to store or validate the attack IP address list, and their blocklist/whitelist status as a transaction in the blockchain, and distribute the added transactions to all the nodes (SDN controller in other autonomous systems) in the blockchain. Ethereum smart contracts were used to store the IP addresses with malicious flag status as a transaction. The DDoS detection/mitigation mechanism was tested in Ethereum testing platform Rapsten testing network and also used Ganache for testing in local blockchain network \cite{Paul2018DeployMedium}. \\

\begin{table*}[!h]
%%% \tablesize{} %% You can specify the fontsize here, e.g., \tablesize{\footnotesize}. If commented out \small will be used.

\captionsetup{justification=centering}
\caption{DDoS mitigation near network using Blockchain\label{DDoSnearNW}}
% \\ \hline

\begin{tabular}{|l|l|l|l|l|}
 \hline
\textbf{Title}	& \textbf{Blockchain} & \textbf{Type} & \textbf{Consensus} & \textbf{Technologies} \\
 \hline 

Yeh et al. \cite{Yeh2020SOChain:Blockchain} & Ethereum & Consortium  & Proof of Work & Smart contracts, Swarm, DOS, Bloom filter  \\ \hline

% &  ISP need to take the sole responsibility for data vesifiers in the proposed approach and selecting the data certifier is challenging.

Yang et al. \cite{Yang2019AServices} & Ethereum	 & Permission & Proof of work & Smart Contract  \\ \hline
% & The approach may not be applicable for attackers using spoofed IP addresses to perform DDoS attacks.

Yeh et al. \cite{Yeh2019ATechnology} &	Ethereum & Consortium & Proof of work & Smart contract, Swarm, Oracle \\ \hline

%  & The spoofed IP addresses causing the DDoS are not possible to block using this approach.
Rodrigues et al. \cite{Rodrigues2017Multi-domainBlockchains} & Ethereum & Public & Proof of Work & Smart Contract, SDN and VNF.  \\ \hline
% & 	The paper intended to leverage IP addresses to block for DDoS mitigation, which may not applicable for Attacker Spoofed  IP addresses.

Burger et al. \cite{BurgerZurich2017CollaborativeBlockchains} & Ethereum & Public & Proof of Work & Smart Contract, Bloom filter  \\ \hline

% & According to the authors work, Ethereum as a general-purpose blockchain is not the ideal technology to run DDoS signaling applications

Rodrigues et al. \cite{Rodrigues2017AContracts}	& Ethereum & Public  &	Proof of Work & SmartContract, SDN \\ \hline
% &  The proposed architecture is not applicable for attacker spoofed IP addresses and the Blockchain transaction capacity will be maximized for covering for all IP address range.

Rodrigues et al. \cite{Rodrigues2017EnablingBloSS} & Ethereum  & Consortium & Proof of Work & Smart Contract, IPFS, SDN \\  \hline
% & The scalability of blockchain for the number of IP addresses(block/white list stored in transactions) is not addressed. 

Hajizadeh et al. \cite{Hajizadeh2020} & Hyperledger Fabric & Private & Kafka & Chain code, SDN,
Threat Platform  \\ \hline
% &	It is not recommended for real production environments due to the absence of fault tolerance

Essaid et al. \cite{Essaid2019ARNN-LSTM} & Ethereum & Public & Proof of work & Smart Contract, Deep learning(LSTM), SDN\\ \hline

%  &	Real-time standard DDoS Datasets are needed to apply the Deep learning methods and spoofed IP address based DDoS attacks are not stoppable with this method. 

Aujla et al. \cite{Aujla2020BlockSDN:Applications} & Generic & Private & - & SDN  \\ \hline

% & The proposed approach required storage capacity in each switch, computational capacity at the controller , the additional cost due to computational needs, interoperability due to switch hardware is not vendor specific.

Shafi et al. \cite{Shafi2019DDoSThings} &	Hyperledger & -	 & Kafka &  SDN, IoT  \\ \hline

% & The approach don’t support the malicious IoT devices in Non SDN network to mitigate the DDoS attacks.

Pavlidis et al. \cite{Pavlidis2020OrchestratingCollaborations} & Ethereum & Public, Private & Proof-of-Authority  & Smart Contract  \\ \hline
 
%  &
% Smart Contract, Programmable Data Path, IPFS & The network level DDoS mitigation may seem too difficult to identify slow DDoS attacks.

Abou et al. \cite{Abou2019Co-IoT:SDN} &	Ethereum & Public & Proof of work & Smart Contract, Software Defined Networking \\ \hline

% & The autonomous system network devices must support SDN and needed controller presence to act as a Blockchain node for Autonomous system level DDoS mitigation. 

%\bottomrule
\end{tabular}
\end{table*}

Yeh et al. \cite{Yeh2020SOChain:Blockchain},  Yeh et al. \cite{Yeh2019ATechnology}, Shafi et al. \cite{Shafi2019DDoSThings} and Hajizadeh et al. \cite{Hajizadeh2020} discussed the threat information sharing including DDoS threat data among the collaborators for secure data sharing using blockchain based smart contracts technology and decentralized data storage. The Security operation centers can be upload the threat data and ISP act as verifier to confirm the illegitimacy of the threat data prior to adding to the blockchain transaction in \cite{Yeh2020SOChain:Blockchain}, \cite{Yeh2019ATechnology}. The Ethereum based smart contract implementation for DDoS data sharing is performed for evaluation. But, in \cite{Hajizadeh2020} and \cite{Shafi2019DDoSThings}, the Hyperledger caliper is used to implement the threat information sharing among the organizations. Each organization may have the SDN controller to run the blockchain application and act as a blockchain node for updating the threat information in other nodes. \\

Rodrigues et al. \cite{Rodrigues2017Multi-domainBlockchains} \cite{Rodrigues2017AContracts} \cite{Rodrigues2017EnablingBloSS} proposed the Ethereum based architecture for DDoS mitigation and their hardware implementation to allow or block the malicious IP addresses in the ISP level. Each transaction may include the IP address and their status to detect the malicious IP address performing the denial of service attacks. The main limitation of the IP address data storage in the transactions may have limitations. But, Burger et al. \cite{BurgerZurich2017CollaborativeBlockchains} discussed that Ethereum is not an ideal technology for DDoS attack IP based signaling using blockchain due to the scalability issue. The authors also mention that Ethereum smart contracts can be applicable for small number of IP addresses space related applications. They recommend that storing the list of IP address in a file storage like IPFS, and the URL of the storage location  can be pointed to the blockchain transactions, and the  location integrity is verified using hash value.  \\

Pavlidis et al. \cite{Pavlidis2020OrchestratingCollaborations} proposed a blockchain based network provider collaboration for DDoS mitigation.  The AS's are selected based on the reputation scores to participate in the DDoS mitigation plan. The programmable data planes are used to implement the mitigation mechanism for DDoS attacks, which is in contrast to most of the works using SDN Openflow protocol. 

\begin{table*}[!h]
%%% \tablesize{} %% You can specify the fontsize here, e.g., \tablesize{\footnotesize}. If commented out \small will be used.
\captionsetup{justification=centering}
\caption{Advantages and limitations of near network based  Blockchain solutions} \label{DDoSusingBC}
% \\ \hline

\begin{tabular}{|l|p{4.5cm}|p{4.5cm}|p{3.4cm}|l|}
 \hline
\textbf{Title} & \textbf{Objective} & \textbf{Advantage} & \textbf{Limitations}\\
 \hline
%\\
% \endfirsthead
% \hline
% \endhead
% \hline
% \multicolumn{3}{r}{continued on next page}\\
% \hline
% \endfoot
% \hline
% \endlastfoot

Yeh et al. \cite{Yeh2020SOChain:Blockchain} & Decentralized DDoS info sharing 
& SOC may use DDoS data among peers &  Selecting the data certifier is challenging \\ \hline
% a  DDoS data exchange platform for security operating center & A novel dual-level Bloom filter developed to protect the privacy of the uploaded and purchased DDoS information. &	ISP need to take the sole responsibility for data verifiers in the proposed approach and selecting the data certifier is challenging. \\ \hline
Yang et al.  \cite{Yang2019AServices} &	Blockchain based DDoS mitigation services & Client validation and provider authentication & Spoofed IP's are ignored  \\ \hline
	
Yeh et al. \cite{Yeh2019ATechnology}  &  Collaborative DDoS info sharing   & SOC info share platform & Spoofed IP's are ignored \\ \hline
% The author's approach is suitable for DDoS data sharing and suffices CIA triad as well as Scalability compared to prior art.	 & The spoofed IP addresses causing the DDoS are not possible to block using the author's approach. \\ \hline

% To proposes a blockchain based online trading system for DDoS mitigation services. &	The approach validate the client and DDoS mitigation service provider authenticity as well as credibility using blockchain to provide DDoS mitigation services. &	The approach may not be applicable for attackers using spoofed IP addresses to perform DDoS attacks \hline

Rodrigues et al. \cite{Rodrigues2017Multi-domainBlockchains} & Blockchain based DDoS mitigation architecture & First architecture for DDoS and Blockchain &  Spoofed IP's are ignored  \\ \hline

% an architecture for multi-domain DDoS Mitigation based on Blockchains, SDN and VNFs. & First time an architecture for multi-domain DDoS Mitigation based on Blockchains and SDN. &	The paper did not provide a proof of concept and intended to  leverage IP addresses to block for DDoS mitigation, which may not applicable for Attacker Spoofed  IP addresses. \\ \hline 

Burger et al. \cite{BurgerZurich2017CollaborativeBlockchains} &	 Scalable Ethereum based DDoS detection  & Practical implementation & Questions on Ethereum usage \\ \hline

% Collaborative  Ethereum based DDoS attacks mitigation scalable approach & Three prototypes of a smart contract for signaling DDoS attacks for the Ethereum developed, tested. &	According to the author's work, Ethereum as a general-purpose blockchain is not the ideal technology to run DDoS signaling applications \\ \hline

Rodrigues et al. \cite{Rodrigues2017AContracts}	 & Blockchain architecture and design for DDoS & Detection and mitigation also included & not for spoofed IP \\ \hline
% of a collaborative mechanism using smart contracts  & The implementation of the Ethereum smart contracts to whitelist/blocklist the IP addresses at the AS level to mitigate DDoS attacks. & The technique is not applicable for attacker spoofed IP addresses and the Blockchain transaction capacity will be maximized for covering for all IP address range.\\ \hline

Rodrigues et al. \cite{Rodrigues2017EnablingBloSS}  & Ethereum testbed for DDoS mitigation & Tested on hardware & Scalability \\ \hline
% a novel approach deploying hardware to simplify the signaling of DDoS attacks &	The authors deployed the hardware to test the Ethereum and SDN based DDoS attack mitigation in a cooperative environment. & The scalability of blockchain for the number of IP addresses(block/whitelist stored in transactions) is not addressed. \\  \hline

Hajizadeh et al. \cite{Hajizadeh2020}  &  Blockchain based threat intelligent platform & Important security application & Fault tolerance \\ \hline

% afor DDoS knowledge sharing across the AS’s &	Permission based blockchain knowledge sharing threat intel platform to share the security information and provided the case study on DDoS attack info sharing. &	It is not recommended for real production environments due to the absence of fault tolerance \\ \hline

% Aujla et al. \cite{Aujla2020BlockSDN:Applications} &  Blockchain as a service to protect the smart city applications deployed in SDN, specific to DDoS. &	A Blockchain based framework to handle various security attacks in Software Defined networks. &	The proposed approach required storage capacity in each switch, computational capacity at the controller, the additional cost due to computational needs, interoperability due to switch hardware is not vendor specific. \\ \hline

Shafi et al. \cite{Shafi2019DDoSThings}	 &  Mitigate the IoT based DDoS attempts in SDN & - & Not support for non-SDN \\ \hline
% a mechanism to  networks using Blockchain  &	Leveraging SDN paradigm to block the IoT attack traffic and using controller as a blockchain node to approve the flow rule approval in the SDN switches. & The approach don’t support the malicious IoT devices in Non SDN network to mitigate the DDoS attacks. \\ \hline

Essaid et al.  \cite{Essaid2019ARNN-LSTM}  & DL and smart contract DDoS detection & DL based & Standard dataset \\ \hline

% a Deep learning-based DDoS attack detection and mitigating the attacks using the Smart contract. & Utilizing the combination of Deep learning(LTSM) and Blockchain technology for handling the DDoS attacks. &	Real-time standard Datasets are needed for applying the Deep learning methods and storing the digital assets data in Blockchain Lightening network. \\ \hline

Pavlidis et al. \cite{Pavlidis2020OrchestratingCollaborations}  & collaborative DDoS mitigation at the AS level & Network level DDoS mitigation & Difficult to identify slow DDoS attacks \\ \hline

% a  schema using blockchain smart contracts 
% & The proposed approach allocates the defenses resource with the help of reputation score and also contains the traffic filtering mechanism. & The network level DDoS mitigation may seem too . 
% % considering the AS reputation score and network flow weight and filter the malicious traffic using programmable data paths.

Abou et al. \cite{AbouElHouda2019Cochain-SC:Contract} &	 Intra-domain and inter-domain DDoS mitigation  & Effective DDoS mitigation & Spoofed IP's are ignored \\ \hline

%  combining SDN, blockchain and smart contract. &	Effectively mitigates the DDoS attack near origin of the attacker  and reduces the cost caused due to passing the network attack traffic along the network traffic. & The autonomous system network devices must support SDN and controller presence to add act the AS as a Blockchain node for DDoS mitigation. \\ \hline
%\bottomrule
\end{tabular}
\end{table*}

In the papers \cite{Manikumar2020BlockchainTechniques} \cite{Essaid2019ARNN-LSTM}, the machine learning algorithms such as K-nearest neighbors (KNN), decision tree and random forest as well as deep learning technique long short-term memory (LSTM) are applied to the network traffic to determine the DDoS attack and considered blockchain technology to whitelist/blocklist the IP addresses at the autonomous system level of the network. But, the machine learning application on the network traffic requires infrastructure and computation capabilities, and ownership responsibility to allocate the resources need to be addressed. Any specific entity like ISP, security service providers will not be interested to perform data analytics unless they have any monetary benefits or business advantages. \\

Overall, we can clearly see that the combination of SDN in AS level and Ethereum smart contract can be implemented to track the IP addresses status and update all the nodes across the internet to mitigate the DDoS attacks. However, there are some limitations like blockchain integration with legacy networks, handling spoofed IP addresses need to be solved for adopting the blockchain based DDoS mitigation in the network level.\\

\subsection{Near attack domain location}
The DDoS attacks mitigation at the attacker network is an effective way to handle DDoS attacks, as the attack traffic will not be propagated to the internet network. Most of the latest DDoS botnets are formed by compromising the legitimate IoT devices located all over the internet and target the victims to send malicious network traffic. So, detection and mitigation of IoT botnets at the source network in essential. Chen et al. \cite{Chen2020ADevices} focused on detecting and mitigating IoT based DDoS attacks or botnets in IoT environment using blockchain. The edge devices or IoT gateways acts as a blockchain node to perform transactions when a network anomaly or attack detected in the IoT environment. The techniques used for network traffic analysis in the paper include statistical analysis, conventional bot detection techniques like community detection. The smart contracts are used to write  attack alerts data in transactions and Ethereum network distribute the data across the IoT nodes. But, the IoT gateway nodes are not usually customer-centric and deploying the blockchain client application in the gateway is challenging for real-time production environment. \\

\begin{table*}[!h]
\captionsetup{justification=centering}
\caption{DDoS mitigation near attack location using Blockchain.\label{DDoSNearattack1}}
\resizebox{\textwidth}{!}{%
\begin{tabular}{|l|p{3.2cm}|l|l|l|l|}

 \hline
\textbf{Title}	& \textbf{Blockchain} & \textbf{Type} & \textbf{Consensus} & \textbf{Technologies}\\
 \hline 

Chen et al.  \cite{Chen2020ADevices} & Ethereum & Public 
& Proof of work &  Smart contract, IOT  \\ \hline
% & The IoT gateway nodes are not usually customer centric and deploying the blockchain client in the gateway is challenging in real time

Javaud et al.  \cite{Javaid2018MitigatingBlockchain} &	Ethereum & Public & Proof of work & Smart Contract, IoT   \\  \hline

% & The server has to use additional resources to validate the IoT devices.The solution is deployed near the victim network and attack traffic consumes the network along the path.

Sagirlar et al. \cite{Sagirlar2018AutoBotCatcher:Things} & Hyperledger 
(Future work) &  permission & BFT & IoT, Chaincode  \\  \hline

Spathoulas et al. \cite{Spathoulas2019CollaborativeBotnets}
& Ethereum (Future work) & Public & Proof of work & IoT, Smart Contract  \\  \hline

% & The network flow as a transaction makes it impractical to add each network flow transactions in the blockchain, as network traffic generates huge amount of traffic.

Abou et al. \cite{Abou2019Co-IoT:SDN} & Ethereum & Permission & Proof of work &  SDN, IOT \\ \hline

% 	& The approach may not be applicable for attackers using spoofed IP addresses to perform DDoS attacks.\\
	
Kataoka et al. \cite{Kataoka2018TrustSDN} & Ethereum & Public, Private & Proof of work & Smart Contract, SDN, IoT  \\  \hline

% & The openflow switch rules need to be updated in this apporach for mitigation and hence the technique not applicable to non sdn networks.

% Kim et al. \cite{Kim2018DDoSBlockchain} & Generic & Private &
% BFT	 & CDN  \\  \hline

% & The approach is mitigating the attacks near the victim network and consume internet  network with attack traffic before mitigation.

% & The IoT gateway are multi-vendor devices and interoperability is an issue.
% The Communication overhead and computation power due to the agents might affect the normal utility of the gateways.

\end{tabular}%
}
\end{table*}

Javaid et al. \cite{Javaid2018MitigatingBlockchain} discussed the blockchain based DDoS attack detection on the servers connected to the IoT devices. The IoT devices sending data to the server is approved by the Ethereum network with an expense of gas cost. When a rogue IoT device trying to send the malicious network traffic, the IoT device is penalized with high gas cost and only trusted devices are approved for connecting to the network. The integration of the IoT with Ethereum enables the denial of service mitigation on the IoT device connected servers. Sagirlar et al. \cite{Sagirlar2018AutoBotCatcher:Things} proposed a blockchain solution for detecting the IoT related peer to peer botnets. The assumption is that botnets frequently communicate to each other to perform malicious activity. The authors mentioned that the network traffic between the botnet nodes are considered as blockchain transactions in permissioned Byzantine Fault Tolerant (BFT) and use these transactions to identify the botnet IoT devices. The proposal method may not be a viable solution, as the network traffic flows are enormous and blockchain may not accommodate the transaction capacity needed for storing in blockchain nodes. \\

Spathoulas et al. \cite{Spathoulas2019CollaborativeBotnets} presented an outbound network traffic sharing among the blockchain enabled IoT gateways to detect the IoT botnet. The authors performed simulations on the proposed solution and showed the promising results using detection efficiency parameter. But, the solution is not tested in the real blockchain nodes installed in the gateway and mentioned that Ethereum smart implementation is one of their future work. But, in general, the IoT gateways are multivendor devices and interoperability among the devices is an issue.\\
 	
Abou et al. \cite{Abou2019Co-IoT:SDN} discussed collaboration among the autonomous systems to detect the DDoS attacks. Each AS contain SDN controller, in which blockchain application like Ethereum client is installed to distribute the malicious IP addresses among other AS's. Whenever a malicious IP address is identified in the AS, the SDN controller updates to the Ethereum client and then Ethereum clients update to all the SDN controller in the AS's for DDoS detection and mitigation. To implement this solution, the AS's should support the same SDN controller and agree to collaboratively work for DDoS mitigation. Kataoka et al. \cite{Kataoka2018TrustSDN} presented a similar \cite{Abou2019Co-IoT:SDN} blockchain and SDN based architecture for whitelisting the IoT devices in the network. The trusted profile consist of IoT devices will be stored in smart contract based blockchain transaction and the SDN controller will update all the switches and routers in the SDN network. This implementation enable the malicious or IoT botnets will be blocked in the attack network itself and protect the networks. Considering there is a huge number of IoT devices connected to internet approximately 31 billion devices as of 2020, the implementation of the blockchain for each gateway in IoT environment is challenging and practically impossible. In addition, the IoT gateway vendors interoperability and supporting the blockchain nodes just for the sake of DDoS detection and mitigation may not seem to be reasonable with the current state-of-the-art technology. \\

% Lo-Yao Yeh et al. \cite{Yeh2019ATechnology} proposed a consortium based blockchain to provide the DDoS data exchange service and share the attack IP address data among Security operation centers(SOC). Ethereum smart contracts are used to implement the blockchain functionality and swarm P2P distributed file storage system to store the list of IPs. The ISP must verify the IP addresses using Decentralized Oracle service. 

\begin{table*}[!h]
\captionsetup{justification=centering}
\caption{Advantages and limitations of near attack location based blockchain solutions} \label{DDoSNearattack}
%\resizebox{\textwidth}{!}{%
\begin{tabular}{|p{2.4cm}|p{5cm}|p{4.5cm}|p{4cm}|}

\hline
\textbf{Title}	& \textbf{Objective} & \textbf{Advantage} & \textbf{Limitations}\\ \hline

Chen et al.  \cite{Chen2020ADevices} & IoT based DDoS detection using blockchain  &	The Attacks can be stopped at the source network & Practically may not be viable	 \\
\hline

Javaid et al. \cite{Javaid2018MitigatingBlockchain} & Ethereum  and IoT integration for DDoS & Automated control of the server IoT inbound traffic & Only applicable to server DDoS  \\ \hline

Sagirlar et al. \cite{Sagirlar2018AutoBotCatcher:Things}  & IoT botnet detection using BFT. &	First blockchain-based IoT botnet detection	& May not be scalable  \\ \hline

Spathoulas et al. \cite{Spathoulas2019CollaborativeBotnets} & IoT botnets detection using blockchain & Outbound traffic exchange using IOT gateway  & Not practically implemented \\ \hline 

Abou et al. \cite{Abou2019Co-IoT:SDN}  & AS level SDN and blockchain solution & Network level DDoS detection & AS legacy networks issue \\ \hline

Kataoka et al. \cite{Kataoka2018TrustSDN}  &  IoT botnets detection using SDN and blockchain & Attacker location based detection & Not applicable to non SDN based IoT   \\ \hline
\end{tabular}

%%}
\end{table*}

\subsection{Near Victim Location}
Yang et al. \cite{Yang2019AServices}  proposed a real-time DDoS mitigation service leveraging a consortium based or permissioned blockchain. Each DDoS service provider has an account in the permission blockchain to provide DDoS mitigation service. The victim looks for the attacker IP-AS mapping in the blockchain, and the trusted service provider IP tagged with AS is authorized to provide the  DDoS mitigation service. The authors also proposed the reputation or credibility validation mechanism of the service providers. However, if the attack IP is spoofed, the author's proposed blockchain based DDoS mitigation service is not applicable. Kyoungmin Kim et al. \cite{Kim2018DDoSBlockchain} proposed a decentralized CDN service to mitigate the DDoS attacks with the help of private blockchain and particularly used by government and military agencies to protect their service. The victims usually the service providers hosting the web content servers. They can protect the servers using the decentralized the CDN services. 

The context of the attacker and victim location may be changed based on the attack type and how the attack is conducted. For example, an attacker may use their infrastructure to send the malicious traffic. In this case, the blockchain based solutions proposed in the attacker domain can be considered as near attacker based solutions. Additionally, the attacker compromise the legitimate IoT devices and use them as a botnet to attack another victim. Here, the solutions deployed in the IoT device locations also comes under near attacker based solutions. The solutions solely implemented in the main victim (not the legitimate IoT bot owner victim) are considered under the Near victim location based solutions. We can say that near the victim based solution  research articles are far too less than the network based and near attacker based solutions. It is too late to mitigate the DDoS attacks near the victim. So, the existing solutions mainly focused on the network level or near attacker.

\subsection{Hybrid solutions}
The hybrid DDoS detection and mitigation solution can be the combination of the network based, near attacker location and the near victim location based solution. For effective mitigation of the DDoS attacks, the multi level mitigation solutions are needed. But, the implementation of these solutions require the collaboration among stakeholders. Abou et al. \cite{AbouElHouda2019Cochain-SC:Contract} proposed intra domain and inter domain DDoS detection and mitigation solution using blockchain. The intra-domain detection include near the victim based solution and inter domain detection meaning that network based solution. The Ethereum smart contract is deployed in each AS to distribute the DDoS threat information and the SDN controller is used to update the AS network traffic filtering rule to block the malicious traffic for inter domain DDoS mitigation. On the other hand, the traffic from switches and routers in the same domains are monitored using SDN controller applications and apply the flow control rules in switches/routers using open flow switch protocol. This mechanism mitigate the internal attacks originating from the same domain. Based on our research, there is limited work done on proposing solutions in multi levels of internet architecture and scope for new research contributions in this area.  

%% file: 6challenges.tex
\section{Open Challenges}
\label{sec:openchallenges}
In this section, we discuss the research challenges to leverage the blockchain technology for DDoS attack detection and mitigation solutions. The detail description of the  decentralized technologies adoption in conventional network issues are presented to handle the DDoS attacks. 

\subsection{Integration with Legacy Network}
Distributed denial of service attacks mitigation involves the network operators, internet service providers and edge network service providers to respond and block the malicious actor traffic. These stakeholders run the network services in legacy platforms and has been providing services for decades and adapting to the decentralized blockchain technology is a major concern. The reasons could be the lack of memory and computation requirements for blockchain in legacy networks \cite{Hajizadeh2020}, trust on the technology, unavailability of blockchain professional workforce, fear of failure to protect customers while using blockchain. In addition, a collaboration between the  ISP’s is required to share the malicious data indicators among the ISP’s and all the stakeholder’s may not be comfortable, as there is no monetization aspect for the internet service providers and usually only benefited by the attack victims. So, a responsible organization or service provider should be stepped up to coordinate among the stakeholders and make sure the involved stakeholders get benefited.

\subsection{Bitcoin/Ethereum P2P Network Zero-Day Vulnerabilities}
The Blockchain transactions process include the network traffic passing through the internet from one node and other nodes in the network; the cryptocurrency exchanges can also act as a blockchain node on behalf of the client and perform the transactions in the exchange conventional network. The attack vector for the blockchain is quite broader and the cost of a single vulnerability in the applications is in millions of dollars. For instance, a parity check vulnerability in Ethereum causes lost \$300 million dollars \cite{2017HowNoon} and a small bug found in cryptocurrencies has a huge impact on the decentralized network. It is also important to note that the  cryptocurrency exchanges having conventional network will have a major consequence to impact the P2P applications. We envision that there is a scope to progress for developing the flawless applications and monitoring the traffic for illegitimate activity detection.

\subsection{Lack of Blockchain P2P Network Datasets}
Monitoring the anomalous behavior of the blockchain network traffic and transactions dataset using machine learning and deep learning techniques is one of the solutions for detecting the DDoS attacks proposed in the prior art \cite{Tayyab2020ICMPv6} \cite{Rathore2019}. But there are very few datasets available in public for continuing research and improving the detection metrics. Mt.Gox exchange trading activity data from 2011 to 2013 is available for public to use for research purpose \cite{VasekMarie2014}. The quality of the data and how older the data is questionable for testing and detecting the real time attacks. We believe that having standard datasets and application of big data analytics in the future is a must requirement for research progress in DDoS detection in cryptocurrency networks.

\subsection{Spoofed IP DDoS Attacks Detection}
The proposed solutions for DDoS attacks detection mainly identifies the source IP address and use blockchain technology to store the transactions and share the IP address among the stakeholders to block/whitelist the IP address with trust and validation at the network level \cite{Abou2019Co-IoT:SDN}\cite{Yeh2019ATechnology}  \cite{Essaid2019ARNN-LSTM} \cite{Yang2019AServices} \cite{Abou2019Co-IoT:SDN} \cite{Kataoka2018TrustSDN} \cite{Rodrigues2017Multi-domainBlockchains} \cite{BurgerZurich2017CollaborativeBlockchains}. These solutions assume that the originating malicious IP addresses are not spoofed, and this condition is not always true. In most of the scenarios, as seen in Table \ref{AttacksHistory}, the attacker performs a reflection attack, in which the spoofed traffic is being sent to the victim to consume the communication capacity or saturating the CPU or memory resources for successful DDoS attack. The researchers also not addressed the IPv6 traffic and can be critical storing the IP version 6 data in blockchain in terms of memory consumption. 

\subsection{IOT and SDN Vendor Interoperability}
The existing state-of-art essentially utilized the software defined networks and internet of things technology to address the denial of service attacks either at the victim level or network level. Even though those solutions prove that the attacks can be mitigated, there is a real challenge when trying to adopt the techniques in industry. The IoT device or gateway vendors are quite diversified and there are multitude of SDN supporting network device providers for enterprise solution. We tend to see incompatibility issue and also supporting blockchain node issues in these network paradigms and deploying a decentralized application across their stakeholder network is impractical. It is desirable to depend on the Blockchain based DDoS mitigation as a service solution like Gladius \cite{2017GladiusMedium}.

%% file: 5futuredirection.tex
\section{Future Directions}
\label{sec:futuredirection}
In this section, the future directions of dealing with DDoS attacks using blockchain technology is explored. We have presented the research directions in terms of the advancements in blockchain and how these advancements can be used to address the DDoS attacks.

\subsection{Internet of Blockchain}
The current blockchain technologies like Bitcoin or Ethereum smart contracts transaction process is sequential and hence, it is very slow to add the  transactions in the blockchain. To solve the scalability  and interoperability issue between blockchain nodes, internet connected blockchain has been proposed and can concurrently process the transactions from  different blockchains. Paralism \cite{2019ParalismBlog}  built the blockchain infrastructure with unlimited scalability and digital economy platform supported by parallel blockchain. Customized script and chain virtualization make paralism support any amount of sub-chains and independently operated chain-based applications and also become the backbone of the internet in decentralized world. This technology is in the early stages of the development and lot of scope to work on utilizing parallel blockchain to share the threat data across the blockchain applications and protect denial of service attacks. We also think that the parallel blockchain surfaces new security issues including leaking the information between the blockchain applications and  will be the topic to focus for researchers while building the blockchain internet backbone. Another notable advancement in the blockchain is Xrouter, which acts as blockchain router to communicate one blockchain like bitcoin to smart contracts, supporting interchain and multichain services \cite{2019IntroducingRouter}.

\subsection{Programmable data planes (P4) for Blockchain based DDoS Solutions}
The network paradigms keep changing as the new technology trends emerged in the enterprises. The Internet of Things supports IP protocol and IoT application  protocols MQTT, XMPP, AMQP etc. The  denial of service attacks can be carried by leveraging the weaknesses in the protocol and flooding the traffic on the victim machine. The combination of Programmable data planes at the gateway level and the blockchain technology for sharing the attack data is effective for mitigation of the attacks. The P4 device in the switch level  that can parse any type of network protocol and makes easy for applying the blockchain technology. We envision that the future work would be proposing new architecture with P4 for mitigation of attacks, developing smart contracts for the gateway level device to monitor and mitigate the attacks using Programmable data planes. 

\subsection{Threat Information Sharing using Blockchain}
Consortium or private based blockchains are most compatible for sharing the threat information among the Blockchain participants. Numerous Ethereum based techniques has applied to share the information with integrity and anonymity. Leveraging the decentralized file storage such as swarm, IPFS enables to store the information rather than keeping the data in transactions and causing time delay to process the sequential transactions. We believe that the information sharing field using blockchain requires improvement and architecture changes to implement secured information sharing network.

\subsection{Ethereum 2.0 Network for DDoS mitigation}
DDoS solutions implemented using Ethereum network \cite{Yeh2019ATechnology} \cite{Abou2019Co-IoT:SDN}faces scalability, speed challenges, in particular transactions refer to allow or block attack IP addresses. Ethereum 2.0 has been proposed and implemented for the last few years \cite{ReneMillman2020WhatDecrypt}. From August 2020, the upgradation to Ethereum 2.0 is initiated with three phases to complete the process. ETH 2.0 works-based proof of stake (POS) rather than POW, which is a major change and the upgradation supports the drastic increase in network bandwidth, Lower Gas Costs and benefit for scalability of the network. We envision implementing the DDoS mitigation scheme in Ethereum 2.0 in the near future. 

%% file: 7conclusion.tex
\section{Conclusion}
\label{sec:conclusion}
Blockchain is emerged as a disruptive technology in recent times and the blockchain application capabilities are promising to use in the field of cybersecurity. DDoS attacks are well known and still considered as a major threat to disrupt the businesses. We have performed a detailed review of the blockchain based solutions for DDoS attacks detection and mitigation including the consideration of the different network environments such as SDN, IoT, cloud or conventional network. The solutions are categorized based on the solution deployment location such as network based, near attack location, near victim location and hybrid solutions. We determined that most of the existing solutions focused on storing the malicious IP addresses in blockchain transactions implemented using smart contract and distribute the IP addresses across the AS's in the network level. However, limited research is performed to propose near victim location and hybrid solutions. Finally, we described the open challenges based on the existing research contributions and the future directions based on the advancements in blockchain technologies like parallel blockchain, Xroute, Ethereum 2.0 to effectively handle the DDoS attacks.

We believe that our review will be a great reference resource for readers and the future researchers interested to pursue the research in the combination of Blockchain and DDoS attacks domain.